\documentclass[conference]{IEEEtran}
\IEEEoverridecommandlockouts
\usepackage{cite}
\usepackage{amsmath, amssymb, amsfonts}
\usepackage{graphicx}
\usepackage{textcomp}
\usepackage{xcolor}
\usepackage{algorithm}
\usepackage{algorithmic}

\def\BibTeX{{\rm B\kern-.05em{\sc i\kern-.025em b}\kern-.08em
    T\kern-.1667em\lower.7ex\hbox{E}\kern-.125emX}}

\begin{document}

\title{Freshness-aware Resource Allocation for Non-orthogonal Wireless-powered IoT Networks}

\author{\IEEEauthorblockN{Yunfeng Chen\IEEEauthorrefmark{1}, Yong Liu\IEEEauthorrefmark{1}\IEEEauthorrefmark{2}, Jinhao Xiao\IEEEauthorrefmark{1}, Qunying Wu\IEEEauthorrefmark{1}, Han Zhang\IEEEauthorrefmark{1}, and Fen Hou\IEEEauthorrefmark{2}}
\IEEEauthorblockA{\IEEEauthorrefmark{1}School of Electronics and Information Engineering, South China Normal University, Foshan, China\\
\IEEEauthorrefmark{2}State Key Laboratory of Internet of Things for Smart City, University of Macau, Macau, China\\
\{2021022347, yliu, 2022024885, 2021022354\}@m.scnu.edu.cn, zhanghan@scnu.edu.cn, fenhou@um.edu.mo}}

\maketitle

\begin{abstract}
This paper investigates a wireless-powered Internet of Things (IoT) network comprising a hybrid access point (HAP) and two devices.
The HAP facilitates downlink wireless energy transfer (WET) for device charging and uplink wireless information transfer (WIT) to collect status updates from the devices.
To keep the information fresh, concurrent WET and WIT are allowed, and orthogonal multiple access (OMA) and non-orthogonal multiple access (NOMA) are adaptively scheduled for WIT.
Consequently, we formulate an expected weighted sum age of information (EWSAoI) minimization problem to adaptively schedule the transmission scheme, choosing from WET, OMA, NOMA, and WET+OMA, and to allocate transmit power.
To address this, we reformulate the problem as a Markov decision process (MDP) and develop an optimal policy based on instantaneous AoI and remaining battery power to determine scheme selection and transmit power allocation. 
Extensive results demonstrate the effectiveness of the proposed policy, and the optimal policy has a distinct decision boundary-switching property, providing valuable insights for practical system design.
\end{abstract}


\section{Introduction}
With the development of 5th generation (5G) networks, a multitude of sensors has been deployed for sensing and monitoring in Internet of Things (IoT) networks. 
Traditional performance metrics, such as throughput or delay, are inadequate to measure the freshness of information \cite{b1}.
A novel metric, called age of information (AoI), shows great potential in characterizing the timeliness of data transmission in state update systems.
The AoI is defined as the time elapsed since the generation time of the latest received status update at the destination side, capturing the freshness of information from the perspective of the destination \cite{b2}.

In earlier research, there was a comprehensive exploration of the performance analysis and optimization of AoI in single-source systems.
In \cite{b2} and \cite{b3}, the AoI under first-come-first-served (FCFS) and last-come-first-served (LCFS) scenarios were investigated, respectively.
The results indicated that managing packets in the cache queue could effectively reduce AoI.
Recent research has shifted focus towards AoI-oriented transmission scheduling for practical multi-source systems.
For a single-hop wireless network, the expected weighted sum AoI (EWSAoI) minimization problem was examined in \cite{b4}, considering throughput constraints. 
In \cite{b5}, the random arrival process of state update packets was taken into account, revealing the effects of different queuing policies on AoI.
These prior works investigated AoI performance and scheduling optimization for orthogonal multiple access (OMA)-based systems, serving only one device over a time-frequency resource block.
However, as the number of devices increases, devices' AoI typically monotonically increases.
Therefore, efficient multiple access technology becomes crucial to reduce AoI in wireless networks with a massive number of devices.

As non-orthogonal multiple access (NOMA) can simultaneously serve multiple devices on a specific resource, enhancing spectral efficiency, it has been recognized as a promising technology for 5G and beyond \cite{b6}.
To mitigate NOMA transmission interference, the successive interference cancellation (SIC) technique is employed for information decoding at the receiver \cite{b7}.
Moreover, some research has delved into the impact of NOMA on AoI. 
In \cite{b8}, NOMA demonstrated superior AoI performance compared to OMA at different packet arrival rates.
An adaptive OMA/NOMA scheduling scheme in downlink networks was investigated in \cite{b9} to minimize the EWSAoI. 
In addition, the adaptive transmission scheme to schedule the OMA, NOMA and collaborative-NOMA was investigated in \cite{b10} for a short-packet communication network.

Furthermore, in IoT networks, the achieved AoI is significantly influenced by the power limitations of IoT devices. 
Therefore, it is crucial to consider power consumption to prolong the network lifetime and achieve minimal AoI. 
Radio frequency (RF)-based energy harvesting technology is deemed a promising solution to offer continuous and sustainable energy supply for IoT devices \cite{b11}.
Given these considerations, integrating RF-based energy harvesting with NOMA technologies becomes a natural choice to ensure energy sustainability and maintain fresh information in IoT networks. 
However, the optimal scheduling of the transmission scheme and allocation of network resources to achieve these goals have not been thoroughly investigated.
These observations serve as the motivation for this work.

In this paper, we investigate a wireless-powered IoT network comprising a Hybrid Access Point (HAP) and two IoT devices. 
The HAP is capable of charging devices in the downlink via WET and collecting status updates in the uplink via WIT.
To ensure information freshness, concurrent WET and WIT are permitted, and OMA and NOMA are adaptively scheduled for WIT. 
Specifically, the HAP can adaptively schedule the transmission scheme from WET, OMA, NOMA, and WET+OMA, while allocating the corresponding transmit power to minimize the EWSAoI.
To achieve this, we reformulate the EWSAoI minimization problem as a Markov Decision Process (MDP) and develop an optimal policy.
This policy, guided by instantaneous AoI and remaining battery power, determines the scheduling of the transmission scheme and the allocation of transmit power. 
Extensive results demonstrate the effectiveness of the proposed policy, and the optimal policy has a distinct decision boundary-switching property, providing valuable insights for practical system design.

\section{SYSTEM MODEL AND PROBLEM FORMULATION}

\subsection{System Model}
We consider a wireless-powered Internet of Things network, which comprises a HAP situated at the network center and two randomly deployed devices (s1 and s2). 
Due to the limited energy of the devices, the HAP is tasked with charging them in the downlink via WET and utilizing WIT to collect status updates from the devices in the uplink.
The time is slotted, with each slot having a duration normalized to unity.
This paper adopts block fading channels, where the channel state remains constant within each time slot but varies across different slots. 
At the beginning of each time slot, devices can generate a new data packet, a concept well-known in the literature as the \textit{generate-at-will} model. 

\subsection{Transmission Scheme}
As only a single antenna is deployed, each device operates in half-duplex mode and is either charged by the HAP or scheduled to upload the status update. 
Multiple access for WIT involves the exploitation of OMA and NOMA techniques. 
In the OMA scheme, the HAP determines which device is scheduled to upload the status update. 
With the NOMA scheme, two devices are scheduled to upload status updates simultaneously with different transmit power, known as power domain NOMA. 
The signals are then decoded at the HAP using the SIC technique.
An essential consideration for NOMA transmission is the allocation of transmit power. 
Additionally, in most existing works, the transmission time is typically divided into two orthogonal parts for WET and WIT, respectively. 
In terms of time utilization, this manner may not be the most efficient.
In this work, concurrent WET and WIT are allowed, enabling the HAP to simultaneously charge one device and collect status updates from another device.
Consequently, all available transmission schemes include WET, OMA, NOMA, and WET+OMA, with the goal of adaptively scheduling these schemes to achieve better AoI performance.

We utilize a binary variable $V_0(t)\in\{0,1\}$ to indicate whether WET is scheduled in time slot $t$, where $V_0(t)=1$ signifies that the HAP is charging the device.
Simultaneously, another binary variable $V_n(t)$ is defined to denote whether device $n$ is scheduled to transmit its status update in time slot~$t$.
$V_n(t)=1$ indicates that device $n$ is scheduled for transmission; otherwise, there is no transmission.
Consequently, the received signal of device $n$ at time slot $t$ can be expressed as
\begin{equation}
\begin{split}
y_n(t)&=[1-V_{n}(t)][V_{0}(t)\sqrt{\bar{P}_H L_0 (d_0/d_n)^{\alpha}} g_n(t) S_0(t)\\
&+ {\textstyle \sum_{m=1}^{2}} V_{m}(t) \sqrt{P_m(t)} h_{m,n}(t) S_m(t)+Z_n(t)],
\end{split}
\end{equation}
where $\frac{L_{0}d_{0}^{\alpha}}{d_{n}^{\alpha}}$ and $g_{n}(t)$ denote the large-scale and small-scale fading of the WET channel between the HAP and device $n$, respectively.
Here, $L_{0}$ is the reference path loss coefficient with a reference distance $d_0$, $\alpha$ is the path loss exponent, and $d_{n}$ is the distance between the HAP and device $n$.
The term $h_{m,n}(t)$ represents the small-scale fading coefficient of the link between device $m$ and device $n$ in time slot $t$ when performing WIT, which follows zero mean and unit variance, \emph{i.e.}, $h_{m,n}(t)\sim \mathcal{CN}(0,1)$.
$S_{0}(t)$ and $S_{m}(t)$ are the transmit signals of the HAP and device $m$, respectively, and $Z_n(t)$ represents the background noise at device $n$.
$\bar{P}_H$ and $P_{m}(t)$ denote the transmit power of the HAP and device $m$, respectively.
In practice, it is crucial for multiple devices to share bandwidth and satisfy the total transmit power constraint to manage inter-cell interference \cite{b12}.
Let's define $\bar{P}_s$ as the maximum transmit power of devices.
The actual transmit power of device $n$ can be expressed as $P_n(t) = \alpha_n(t)\bar{P}s$, where ${\alpha}_{n}(t)$ is the power allocation factor of device $n$ in time slot $t$ and must satisfy $0 \le \alpha_n(t) \le 1$. 
When both devices are scheduled to upload status updates, namely NOMA transmission, we have $\alpha_1(t) + \alpha_2(t) = 1$.
Particularly, when $\alpha_n(t) = 1$, it means that device $n$ is scheduled under the OMA scheme with the maximum power $\bar{P}_s$ for WIT.

During the energy harvesting process, the influence of other devices may be disregarded, given the significantly higher transmit power of the HAP compared to that of IoT devices.
Consequently, the energy harvested by device $n$ in time slot $t$ can be expressed as
\begin{equation}
e_n(t)=[1-V_n(t)]V_{0}\left ( t \right )\eta \tau \bar{P}_H L_0 (d_0/d_n)^{\alpha}|g_n(t)|^2,
\end{equation}
where $\eta$ denotes the energy efficiency, and $\tau$ represents the duration of each time slot. 
In this work, the energy harvested by the devices is allocated for WIT, with no consideration given to the energy consumed by the devices during information generation. 
The initial energy of the devices' batteries is defined as the maximum capacity $\mathbf{E}_{max}$, and $\mathbf{E}_n(t)$ signifies the energy that device $n$ possesses at the commencement of time slot $t$, adhering to the constraint $0 \le \mathbf{E}_n(t) \le \mathbf{E}_{max}$. 
Thus, the evolution of device $n$'s energy can be described as
\begin{equation}
\mathbf{E}_n(t)=\mathbf{E}_n(t-1)+ e_n(t-1)-V_n(t-1) \tau P_n(t-1).
\end{equation}

Similarly, the received signal at the HAP can be written as 
\begin{equation}
\begin{split}
y_0(t) &= {\textstyle \sum_{n=1}^{2}}V_n(t) \sqrt{P_n(t) L_0 (d_0/d_n)^\alpha} h_n(t) S_n(t)\\
&+V_{0}(t)\sqrt{\bar{P}_H} h_I(t) S_0(t)+Z_0(t),
\end{split}
\end{equation}
where $h_n(t)$ denotes the small-scale fading of the WIT channel between the HAP and device $n$, and $h_{I}$ represents the self-interference fading channel.
$Z_0(t)$ is the complex additive Gaussian noise with variance $\sigma^{2}$.
To enhance conciseness, we introduce the notation ${\omega}_n(t)=L_0(d_0/d_n){^\alpha}\left|{h}_n(t)\right|^2$ and ${\omega}_0(t)=|{h}_I(t)|^2$ for the relevant random variables.
In addition, $\lambda_n=\frac{1}{E\left(\omega_n\right)}$ and $\lambda_0=\frac{1}{E\left(\omega_0\right)}$ denote the reciprocals of their mean values, respectively.
Given the exponentially distributed nature of the random variable $\omega$ and the parameter $\lambda$, the probability density function of $\omega$ is expressed as $\mathit{f}_{\omega}(\mathit{x})=\lambda e^{-\lambda \mathit{x}}$.
In the Rayleigh fading environment, where there is uncorrelated channel fading of signals from different devices with different path losses, independent non-identical distributions are integral to the proposed system.
Further, we define ${\Phi}_0(t)=\bar{P}_H{\omega}_0(t)$ as the received power of the self-interference signal at the HAP, and ${\Phi}_n(t)={P}_n(t){\omega}_n(t)$ as the instantaneous received power of device $n$'s signal at the HAP.
Consequently, the received signal in Equation (4) can be reformulated as
\begin{equation}
y_0(t)= {\textstyle \sum_{n=0}^{2}} V_n(t) \sqrt{\Phi_n(t)} S_n(t)+Z_0(t).
\end{equation}

In this work, a dynamically ordered SIC decoding scheme is employed, as detailed in \cite{b7}.
Before decoding the device signal, the HAP will order the received instantaneous signal power $\Phi_n(t)$ to determine the decoding order.
The decoding order is denoted by an arrangement A.
For a network bandwidth of 1 Hz, the achievable transmission rate for the n-th decoded device is determined by the expression
\begin{equation}
\begin{split}
&{R}_{A_n}(t)=\\
&\log_2\left(1+\frac{\Phi_{A_n}(t)}{\sum_{m=n+1}^{2}V_{A_m}(t)\Phi _{A_m}(t)+V_{0}\left ( t \right )\Phi _{0}(t)+\sigma^2}\right).
\end{split}
\end{equation}

Moreover, we define $\bar{R}_{n}$ as the target rate of the device $n$.
Since the status update packets have the same length, the target rate of each device is equivalent, i.e., $\bar{R}_{1}=\bar{R}_{2}=\bar{R}$, and the signal-to-noise ratio to guarantee the target rate is ${\beta}=2^{\bar{R}}-1$.

\subsection{Age of information}
We employ the AoI metric to assess the timeliness of status information updates from devices. 
It is defined as the time elapsed since the generation time of the latest received status update at the destination side.
The AoI of device $n$ at time t can be expressed as $\Delta_{n} \left ( t \right ) = t-u_{n}\left(t\right)$, where $u_{n}\left(t\right)$ represents the generation time of the latest received status update packet at time t.
If the HAP successfully receives the status update packet from device $n$ within a time slot, the AoI of device $n$ will decrease to 1; otherwise, its AoI will increase by 1.
Thus, the evolution of $\Delta_{n} \left ( t \right )$ can be expressed as
\begin{equation}
\Delta_n(t+1)=\left\{\begin{array}{cl}
1, & if \ q_n(t)=1,\\
\Delta_n(t)+1, &  if \ q_n(t)=0,
\end{array}\right.
\end{equation}
where $ q_{n} (t) = 1 $ indicates that the HAP successfully received the status update packet from device $n$, otherwise $ q_{n} \left ( t \right ) = 0$. 

Subsequently, the EWSAoI of devices is used to measure the information freshness of the network, which is given by
\begin{equation}
\bar{\Delta}=\lim _{T \rightarrow \infty} \frac{1}{T} \mathbb{E}\left[ {\textstyle \sum_{n=1}^{2}}  {\textstyle \sum_{t=1}^{T}} W_{n} \Delta_n(t)\right],
\end{equation}
where T is the time range of the discrete-time system, and $W_{n}$ is the weight coefficient of device $n$ satisfying $W_{1}+W_{2}=1$. The expectation operator $\mathbb{E}\left [ \cdot \right ]$ considers all possible dynamics. 

\subsection{Outage probability under different transmission scheme}
Based on Equation (7), the evolution of AoI is notably contingent on whether the transmission is successful or not.
Thus, it becomes imperative to derive the transmission success probability.
To tackle this, we analyze the outage probability, which is the probability of failed transmission.

In the WET scheme, the HAP charges both devices, and no device transmits information.
Consequently, the outage probabilities for both devices are equal to 1, \emph{i.e.,} $\mathbb{P}_{1}^{W} = \mathbb{P}_{2}^{W} =1$.

In the OMA scheme, when device $n$ is scheduled for WIT in time slot $t$, the received signal at the HAP, as expressed in Equation (5), is given by 
\begin{equation}
y_{0,n}^{O}(t)=\sqrt{\Phi _n(t)}S_n(t)+Z_0(t).
\end{equation}
Consequently, the outage probability\footnote{The detailed deriving procedure can be referred to the cited literature [7].} representing the probability of the transmission failure for device $n$ is given by
\begin{equation}
\begin{aligned}
\mathbb{P}_n^{O}(t)=1-e^{-\frac{\lambda_n\beta\sigma^2}{P_n}}.
\end{aligned}
\end{equation}

In the NOMA scheme, both devices simultaneously transmit their information. The received signal at the HAP is given by
\begin{equation}
y_{0}^N(t)=\sqrt{\Phi_1(t)} S_1(t)+\sqrt{\Phi_2(t)} S_2(t)+Z_0(t).
\end{equation}

The instantaneous received power of s1's and s2's signals are ranked at the HAP and the order can be expressed as $\Phi_{A_1}>\Phi_{A_2}$, so the decoding order is $\left[\mathrm{A}_1,\mathrm{A}_2\right]$. According to the SIC decoding principle, HAP first decodes the signal from device $\mathrm{A}_1$ while treating the signal from device $\mathrm{A}_2$ as interference. Therefore, the outage probability of device $\mathrm{A}_1$  is
\begin{equation}
\begin{aligned}
\mathbb{P}_{A_1}^{N}(t)=1-\frac{\lambda_{A_2}P_{A_1}}{\lambda_{A_2}P_{A_1}+\lambda_{A_1}P_{A_2} \beta} e^{-\frac{\lambda_{A_1} \beta \sigma^2}{P_{A_1}}}.
\end{aligned}
\end{equation}

Once the signal of device $\mathrm{A}_1$ is successfully decoded, $\mathrm{HAP}$ extracts it and then decodes the signal from device $\mathrm{A}_2$ without interference. Therefore, the outage probability of device $\mathrm{A}_2$ is 
\begin{equation}
\begin{split}
\begin{aligned}
\mathbb{P}_{A_2}^{N}(t)=&1-\frac{\lambda_{A_2}P_{A_1}}{\lambda_{A_2}P_{A_1}+\lambda_{A_1}P_{A_2} \beta} \\
&\times e^{-\frac{\left ( \lambda_{A_2}P_{A_1}+\lambda_{A_1}P_{A_2}+\lambda_{A_1}P_{A_2}\beta \right)\beta \sigma^2 }{P_{A_1}P_{A_2}}}.
\end{aligned}
\end{split}
\end{equation}

In the WET+OMA scheme, the WET and OMA are enabled simultaneously, and device $n$ is selected for information transfer while the HAP concurrently charges another device. Thus, the received signal of the HAP is
\begin{equation}
y_{0,n}^{W+O}(t)=\sqrt{\Phi_n(t)}S_n(t)+\sqrt{\Phi_0(t)}S_0(t)+Z_0(t).
\end{equation}
Since the HAP can directly decode the signal of device $n$ with self-interference, the outage probability of device $n$ is
\begin{equation}
\begin{aligned}
\mathbb{P}_n^{W+O}(t)=1-\frac{\lambda_{0}P_n}{\lambda_{0}P_n+\lambda _{n}\beta \bar{P}_H} e^{\frac{-\lambda_n\beta\sigma^2}{P_n}}.
\end{aligned}
\end{equation}

It is noteworthy that when the OMA scheme is employed for WIT, the outage probability of the unscheduled device consistently equals 1. 
Additionally, for a more intuitive representation, $\mathbb{P}_{1}^{N}$ and $\mathbb{P}_{2}^{N}$ will be used directly in the following to denote the outage probability of decoding device 1 and device 2 information when using the NOMA scheme, respectively.

\subsection{Problem Formulation}

The goal of this paper is to find the optimal scheduling policy, including adaptively transmission scheme selection of WET, OMA, NOMA, and WET+OMA, and allocate the transmit power to minimize network EWSAoI. Thus, the following optimization problem is formulated as
\begin{equation}
\begin{array}{ll}
\textbf{P1:}\min _\pi  & \bar{\Delta} \\
 \qquad \ \text {s.t. } 
&W_{1}+W_{2}=1,\\
&0 \le \alpha_{n}(t) \le 1,\\
&0 \le \mathbf{E}_{n}(t) \le \mathbf{E}_{max},\\
& \sum_{n=1}^2 \alpha_{n}(t)=1,if \; any \; V_n(t) \ne 0,
\end{array}
\end{equation}
where the policy $\pi$ denotes the sequential actions, including the transmission policy and the transmit power allocation.

\section{Scheduling Policy based on MDP}
From the previous section, we can see that concurrent WET and WIT enable devices to harvest more energy for sustainable operation, while the cost is external interference to WIT, which increases the transmission fail probability. Meanwhile, NOMA provides more opportunities for devices to upload status updates to the HAP. To maintain the timeliness of the status updates uploaded by devices, the HAP must carefully decide which transmission scheme to schedule based on the network state.

To reduce the computational complexity, the transmit power and battery level of devices are considered discrete. Specifically, the transmit power $P_{n}$ can only be taken from the discrete feasible set $\left\{0,p,2p,... ,Lp\right\}$, where $p=\bar{P}_{s} / L$. Thus, the power allocation coefficient $\alpha_{n}$ can only be taken from the discrete set $\left \{0,\frac{1}{L},\frac{2}{L},... ,1 \right\}$. In addition, the condition $\alpha_{1}+\alpha_{2}=1$ must be satisfied, except for the WET scheme. Besides, the battery level $E_{n}(t)$ of device $n$ in time slot $t$ can only be taken from the discrete set $\left\{0,k,2k,... ,Mk\right\}$, where $k=\mathbf{E}_{max}/M$. Therefore, $E_{n}(t)=\left \lfloor \mathbf{E}_{n}(t)/k \right \rfloor$.

The policy $\pi$ consists of a finite sequence of actions over time slots, denoted by $\left \{ a_{t} \right \} =\left \{[V_{0}(t),\alpha_{1}(t),\alpha_{2}(t)] \right \}$, consisting of the transmission schemes and the transmit power allocation. We further use the idea of coding to encode these actions. If $a_{t}=[1,0,0]$ means that the HAP adopts the WET scheme; if $a_{t}\in \left\{\left[0,1,0\right],\left[0,0,1\right]\right\}$ means that the HAP adopts the OMA scheme; if $a_{t}\in \left\{\left[0,\frac{1}{L},\frac{L-1}{L}\right],...,\left[0,\frac{L-1}{L},\frac{1}{L}\right]\right\}$ means that the HAP adopts the NOMA scheme; if $a_{t}\in \left\{\left[1,1,0\right],\left[1,0,1\right]\right\}$ means that the HAP adopts the WET+OMA scheme. 

Through the above analysis, we then investigate the optimal policy to minimize the EWSAoI of the system by formulating the problem (16) as an MDP problem. The MDP problem can be described by a tuple $\left \{ \mathcal{S},\mathcal{A},\mathcal{P},\mathcal{R}\right\} $:

$(1)$ \textbf{State space} $\mathcal{S}\triangleq\left \{\Delta_{1},\Delta_{2}, E_{1}, E_{2}\right\}$: The system state in time slot $t$ consists of the instantaneous AoI and the battery level of both devices. Noting that the size of the MDP state space is infinite because there is no theoretical upper bound on the AoI. Therefore, we consider the truncation case and choose $\Delta_{max}$ as the maximum AoI, \emph{i.e.,} if the AoI increases beyond $\Delta_{max}$, we will reset it to $\Delta_{max}$.

$(2)$ \textbf{Action space} $\mathcal{A}=\left\{a_{t}\right\}$: A detailed description of the action $a_{t}\in\mathcal{A}$ is provided above. It is important to note that, if the device transmits power of the selected action exceeds the maximum transmit power that the battery level of that device can support, the action should be considered invalid and removed from the current action space.

$(3)$ \textbf{Transition probabilities} $\mathcal{P}$: $\mathcal{P}\left(s_{t+1} \mid s_{t},a_{t}\right)$ is the probability moving to the next state $s_{t+1}$ after taking action $a_{t}$ in the current state $s_{t}$. Since the transfer of the energy state is deterministic, we will directly use $E^{'}$ to denote the next state of $E$. Based on the outage probabilities in section II, we have the following transfer probabilities.

$\bullet$ In the case of the WET scheme:
\begin{equation}
\mathcal{P}(\Delta_1+1,\Delta_2+1,E_{1}^{'},E_{2}^{'}\mid s_{t},a_{t} = [1,0,0]) = 1.
\end{equation}

$\bullet$ In the case of the OMA scheme:
\begin{equation}
\begin{array}{c}
\mathcal{P}(1,\Delta_2+1,E_{1}^{'},E_{2}^{'}\mid s_{t},a_{t} = [0,1,0]) = 1-\mathbb{P}_1^{O},\\
\mathcal{P}(\Delta_1+1,\Delta_2+1,E_{1}^{'},E_{2}^{'}\mid s_{t},a_{t} = [0,1,0]) = \mathbb{P}_1^{O},\\
\mathcal{P}(\Delta_1+1,1,E_{1}^{'},E_{2}^{'}\mid s_{t},a_{t} = [0,0,1]) = 1-\mathbb{P}_2^{O},\\
\mathcal{P}(\Delta_1+1,\Delta_2+1,E_{1}^{'},E_{2}^{'}\mid s_{t},a_{t} = [0,0,1]) = \mathbb{P}_2^{O}.\\
\end{array}
\end{equation}

$\bullet$ In the case of the NOMA scheme:
\begin{equation}
\begin{array}{c}
\mathcal{P}((1,1,E_{1}^{'},E_{2}^{'})\mid s_{t},a_{t}) = (1-\mathbb{P}_1^{N})(1-\mathbb{P}_2^{N}),\\
\mathcal{P}((1,\Delta_2+1,E_{1}^{'},E_{2}^{'})\mid s_{t},a_{t}) = (1-\mathbb{P}_1^{N})\mathbb{P}_2^{N},\\
\mathcal{P}((\Delta_1+1,1,E_{1}^{'},E_{2}^{'})\mid s_{t},a_{t}) = \mathbb{P}_1^{N}(1-\mathbb{P}_2^{N}),\\
\mathcal{P}((\Delta_1+1,\Delta_2+1,E_{1}^{'},E_{2}^{'})\mid s_{t},a_{t}) = \mathbb{P}_1^{N}\mathbb{P}_2^{N}.\\
\end{array}
\end{equation}

$\bullet$ In the case of the WET+OMA scheme:
\begin{equation}
\begin{array}{c}
\mathcal{P}(1,\Delta_2+1,E_{1}^{'},E_{2}^{'}\mid s_{t},a_{t} = [1,1,0]) = 1-\mathbb{P}_1^{W+O},\\
\mathcal{P}(\Delta_1+1,\Delta_2+1,E_{1}^{'},E_{2}^{'}\mid s_{t},a_{t} = [1,1,0]) = \mathbb{P}_1^{W+O},\\
\mathcal{P}(\Delta_1+1,1,E_{1}^{'},E_{2}^{'}\mid s_{t},a_{t} = [1,0,1]) = 1-\mathbb{P}_2^{W+O},\\
\mathcal{P}(\Delta_1+1,\Delta_2+1,E_{1}^{'},E_{2}^{'}\mid s_{t},a_{t} = [1,0,1]) = \mathbb{P}_2^{W+O}.\\
\end{array}
\end{equation}
The time subscripts of $\Delta_{nt}$ and $E_{nt}$ are omitted for brevity.
 
$(4)$ \textbf{Reward \enspace function} $\mathcal{R}$: let $\mathcal{R}$ be the one-stage reward function for state $s_{t}$ performing action $a_{t}$, which is defined as
\begin{equation}
\mathcal{R}(s_{t},a_{t})=W_{1}\Delta_{1t}+W_{2}\Delta_{2t}. 
\end{equation}

Given any initial state $s_{0}$, the infinite horizon averaged reward of the policy $\pi$ can be expressed as
\begin{equation}
C(\pi,s_{0}) = \lim_{T \to \infty} sup \enspace  \frac{1}{T}  {\textstyle \sum_{t = 0}^{T}} \mathbb{E}_{s_{0}}^{\pi} \left [\mathcal{R}(s_{t},a_{t})\right ].
\end{equation}

To this end, the problem (16) can be reformulated as the following MDP problem, \emph{e.g.,}
\begin{equation}
\textbf{P2:} \ \underset{\pi}{min} \ C(\pi,s_{0}).
\end{equation}

The optimal policy $\pi^{\ast}$ to minimize the total average cost can be obtained by solving the following Bellman equation~\cite{b13}:
\begin{equation}
\theta^{\ast} + V(s)=\underset{a \in \mathcal{A}}{min}\left \{\mathcal{R}(s,a)+\mathbb{E}\left [V(s')\mid s,a  \right ]\right\},\forall s\in \mathcal{S},   
\end{equation}
where $\theta^{\ast}$ is the optimal average reward of the problem $\left ( 23 \right )$, the value function $V\left ( s \right )$ is the mapping from state $s$ to the reward set $\mathcal{R}$ that satisfies the optimal average reward function, and $s'$ is the next state after state $s$ takes action $a$. 

\vspace{-5 mm}
\begin{figure*}[htbp]
 \setlength{\abovecaptionskip}{-10pt}
 \setlength{\belowcaptionskip}{-10pt}
 \centering
 \begin{minipage}{0.49\linewidth}
  \centering
  \includegraphics[width=0.7\linewidth]{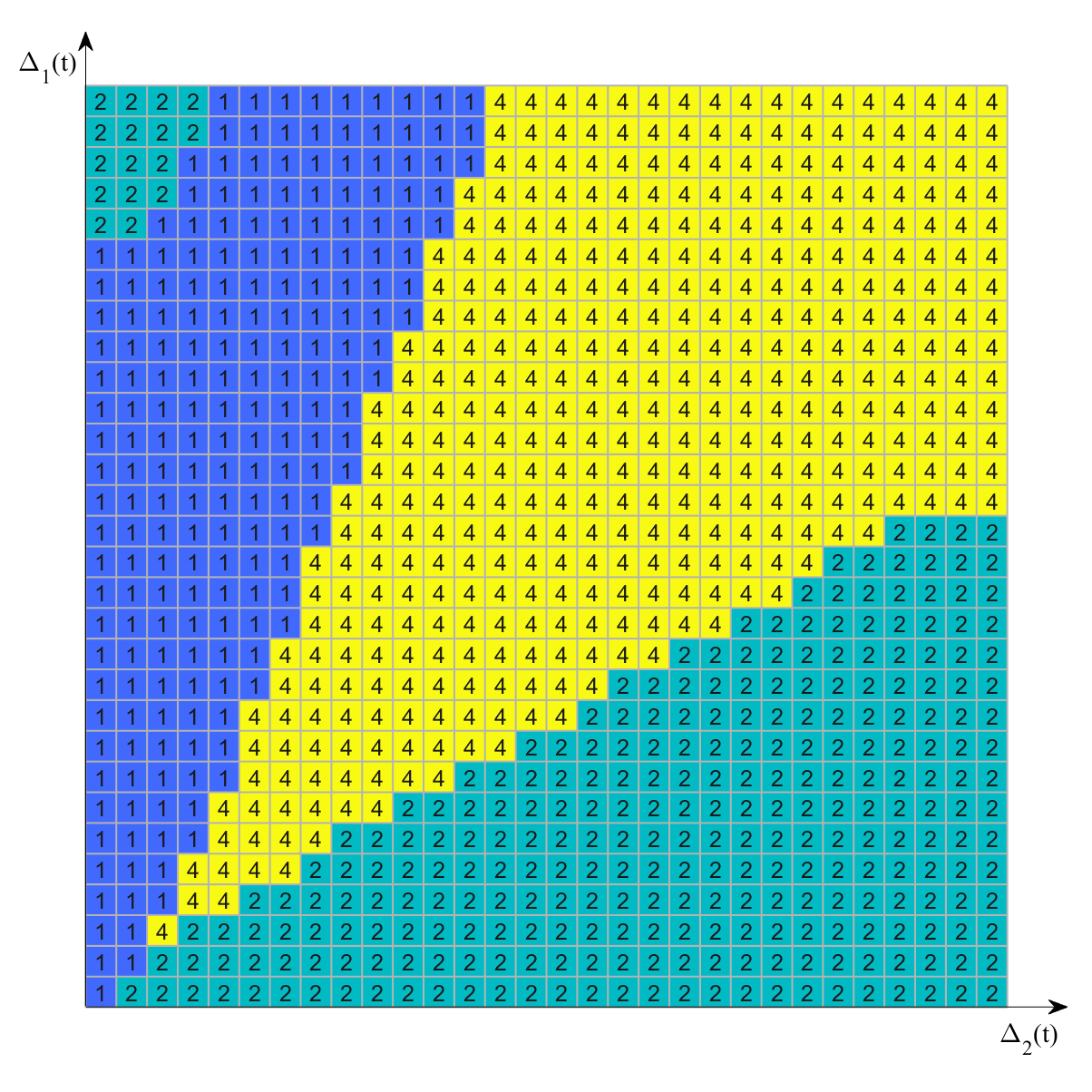}
  \caption{The optimal policy when transmission SNR $\rho$ = 50dB.}
 \end{minipage}
 \centering
 \begin{minipage}{0.49\linewidth}
  \centering
  \includegraphics[width=0.7\linewidth]{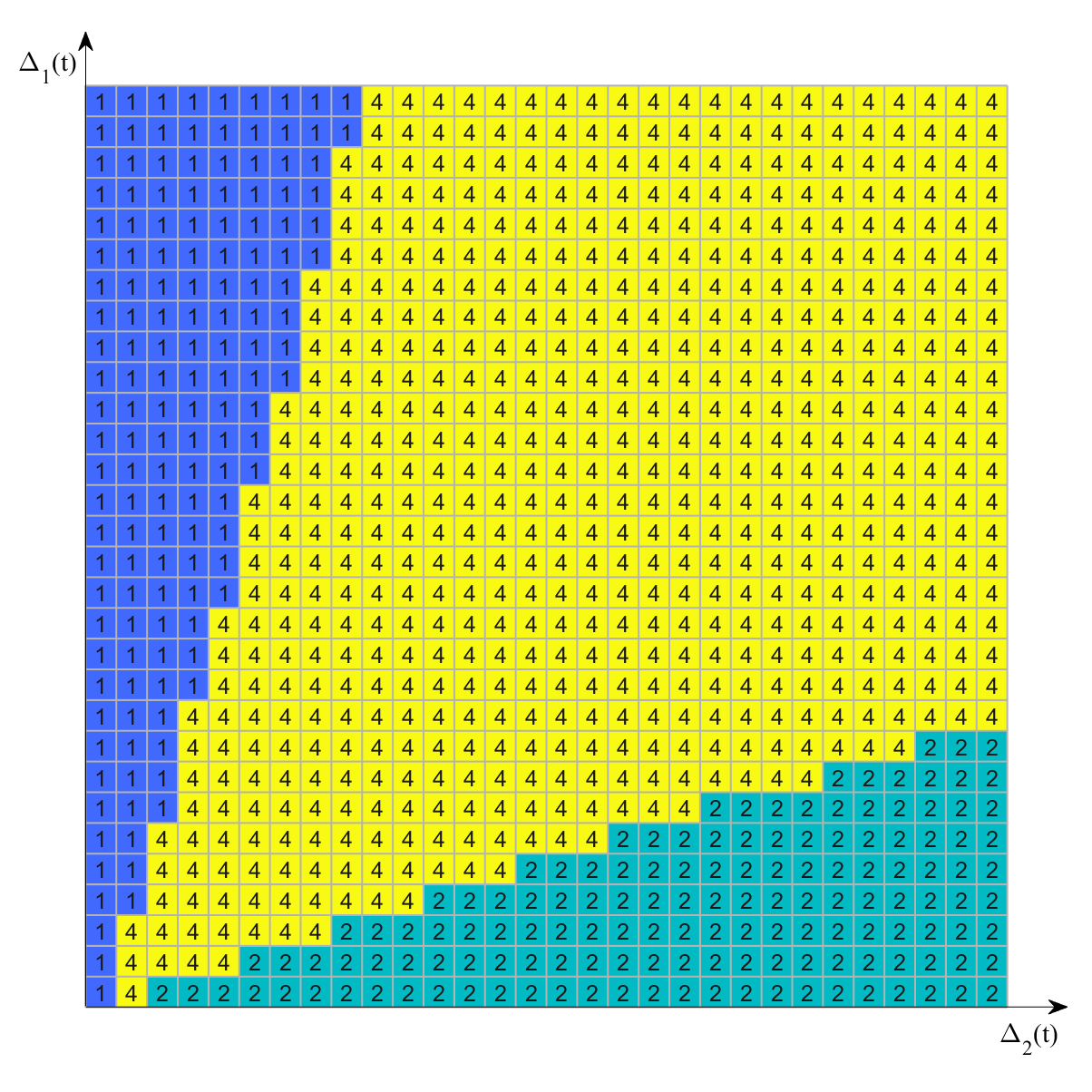}
  \caption{The optimal policy when transmission SNR $\rho$ = 60dB.}
 \end{minipage}
\end{figure*}

To make the equation $\left ( 24 \right )$ converge faster, we introduce a discount factor $\gamma$ and $\gamma \in \left (0,1\right)$. Note that when $\gamma \to 1$, the infinite sum of discounted rewards converges to the expected average return~\cite{b13}. Thus, we can obtain:
\begin{equation}
\theta^{\ast} + V(s)=\underset{a \in \mathcal{A}}{min}\left \{\mathcal{R}(s,a)+\gamma \mathbb{E}\left [V(s') \mid s,a \right ]\right\},\forall s\in \mathcal{S}.  
\end{equation}

Therefore, for any $s\in \mathcal{S}$, the optimal policy $\pi^{\ast}$ can be obtained by the following equation:
\begin{equation}
\pi^{\ast}= \underset{a \in \mathcal{A}}{argmin}\left \{\mathcal{R}(s,a)+\gamma \sum_{s'\in \mathcal{S}}^{} \mathcal{P}\left (s'\mid s,a\right) V\left ( s'\right)\right\}.  
\end{equation}

We apply the policy iteration algorithm to estimate $\pi \approx \pi^{\ast}$.

\begin{algorithm}
  \caption{Policy Iteration Algorithm for estimating $\pi \approx \pi^{\ast}$}
  \label{alg1}
  \begin{algorithmic}[1]
  \STATE{\bf{Input}}
  \STATE \quad All required parameters are shown in section IV.
  \STATE {\bf{Initialization}} 
  \STATE \quad $V\left ( s \right ) \gets 0$ and $\pi\left ( s \right ) \in \mathcal{A}$ arbitrarily for all $s\in \mathcal{S}$.
  \STATE {\bf{Policy Evaluation}} 
  \STATE \quad Loop:
  \STATE \qquad $\varepsilon \gets 0$
  \STATE \qquad Loop for each $s\in \mathcal{S}$:
  \STATE \quad \qquad $v \gets V\left ( s \right )$
  \STATE \quad \qquad $V( s ) \gets \mathcal{R}(s,\pi(s))+\gamma \sum_{s'}^{}\mathcal{P} (s'\mid s,\pi(s)) V ( s')$
  \STATE \quad \qquad $\varepsilon \gets \max ( \varepsilon,|v-V(s)|)$
  \STATE \quad until $\varepsilon < \varepsilon^{\ast}$ (a small positive number determining the
  \STATE \quad accuracy of estimation)
  \STATE {\bf{Policy Improvement}}
  \STATE \quad policy-stable $\gets$ true
  \STATE \quad For each $s\in \mathcal{S}$:
  \STATE \qquad old-action $\gets$ $\pi(s)$
  \STATE \qquad $\pi(s) \gets arg \min_{a} \mathcal{R}(s,a)+\gamma \sum_{s'}^{}\mathcal{P} (s'\mid s,a) V ( s')$
  \STATE \qquad If old-action $\ne \pi(s)$, then policy-stable $\gets$ false.
  \STATE \quad If policy-stable, then stop and return $V \approx v^{\ast}$ and return
  \STATE \quad $\pi \approx \pi^{\ast}$; else go to step 5.
  \end{algorithmic}
\end{algorithm}

\section{NUMERICAL RESULTS}
In this section, we evaluate the proposed transmission scheme and scheduling policy via simulation results. 
Since the communication link environment between HAP and device is similar regardless of whether it chooses WET or WIT,  we consider channel reciprocal, and $h_{n}(t)=g_{n}(t)$. Let the transmit power of the HAP $\bar{P}_H=10W$, the maximum power $\bar{P}_s=0.01$W and $L=10$. The target rate $\bar{R}=2 \enspace bit/s$, the duration of each time slot $\tau = 1s$, the energy conversion efficiency $\eta = 0.5$, the discount factor $\gamma = 0.8$, the parameters $\lambda_{0}=10^8$, $\lambda_{1}=250$, $\lambda_{2}=500$, $\varepsilon^{\ast} = 10^{-4} $ and $W_{1}=W_{2}=0.5$. SNR refers to the signal-to-noise ratio $\rho=\bar{P}_s / \sigma^{2}$ of wireless information transmission. 

We set the maximum battery capacity $\mathbf{E_{max}}=0.02J$, $M=20$ and $\Delta_{max}=30$ so that the truncated finite states basically satisfy the system requirements and thus approximate the countably infinite state space \cite{b14}. The optimal policy when the current state is $s_{t}=(\Delta_1,\Delta_2,11,11)$ at SNR 50 dB and 60 dB are shown in Fig. 1 and Fig. 2, respectively. The number 1 (blue) represents the WET+OMA scheme, the number 2 (green) represents the OMA scheme, the number 3 (pink) represents the WET scheme and the number 4 (yellow) represents the NOMA scheme, indicated by different colors.

The optimal adaptive transmission policy under different SNR of 50dB and 60dB is illustrated in Fig. 1 and Fig. 2, respectively.
When the AoI of s1 is substantial, the HAP tends to opt for the WET+OMA scheme to update the status of s1 while concurrently charging s2.
Since the channel quality of s2 is poor, the energy collected by WET is usually less than s1, so it needs to be charged as much as possible to ensure stable operation.
However, if the AoI of s1 is very large, the HAP tends to adopt the OMA scheme to schedule s1 for state update to avoid an unrestrained increase in the AoI of s1 due to self-interference generated by the energy transmission, which affects the overall performance of the network. If the AoI of s2 is large, the HAP tends to use the OMA scheme to update the status of s2 and does not choose to charge s1 at this time. The reason is that the channel quality of s2 is poor, so the HAP will choose the OMA scheme with the lowest outage probability to schedule device 2 to upload updated status. When the AoI of both devices is relatively large and close, HAP tends to choose the NOMA scheme because NOMA can achieve the simultaneous decrease of the AoI of both devices. Moreover, by comparing Fig. 1 and Fig. 2, we can find that HAP tends to use the NOMA scheme when the SNR is larger and the optimal policy has a precise nature of decision boundary-switching.

\vspace{-4 mm}
\begin{figure}[htbp]
 \setlength{\abovecaptionskip}{-5pt}
 \setlength{\belowcaptionskip}{-5pt}
 \begin{minipage}{1.0\linewidth}
  \centering
  \includegraphics[width=0.85 \linewidth]{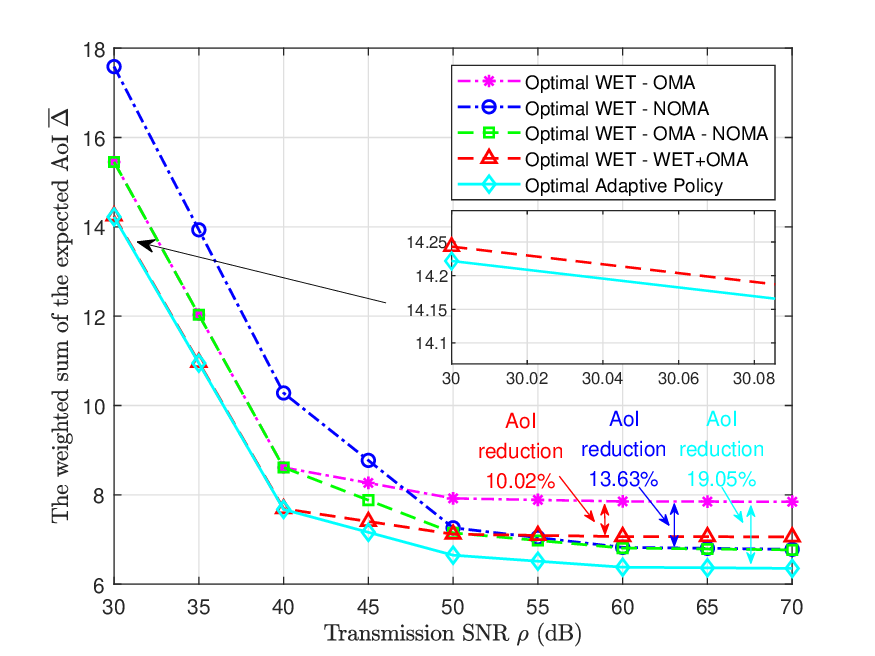}
  \caption{The performance comparison of different policies versus SNR.}
 \end{minipage}
\end{figure}

Fig. 3 displays the EWSAoI performance under different schemes when the current state is $s_{t}=(1,1,21,21)$. 
It is observed that the WET-WET+OMA scheme consistently outperforms the WET-OMA scheme. 
This effectiveness is attributed to the concurrent usage of WET and WIT, effectively alleviating the energy limitations of the devices.
At low SNR, OMA outperforms NOMA because NOMA treats the signal from the other device as interference during the decoding process, leading to a higher outage probability. 
In contrast, at relatively high SNR, the NOMA scheme surpasses OMA and WET+OMA.
This is because the NOMA scheme can concurrently update the status of multiple devices with a low outage probability when the channel quality is better, highlighting the advantage of NOMA in timely status updates.
Additionally, the WET-OMA-NOMA scheme demonstrates the advantages of allowing adaptive scheduling of OMA and NOMA while performing WIT. 
Furthermore, the optimal adaptive scheme performs slightly better than the optimal WET-WET+OMA scheme when the SNR is small.
This is because the optimal adaptive scheme chooses the OMA scheme when the battery is sufficient, avoiding the self-interference caused by the WET+OMA scheme during energy transmission.
In conclusion, our proposed optimal adaptive scheduling scheme consistently outperforms other transmission schemes.

\section{CONCLUSIONS}
In this paper, we addressed the EWSAoI minimization problem in a wireless-powered IoT network. 
To ensure information freshness, we permitted concurrent WET and WIT, with OMA and NOMA being adaptively scheduled for WIT. 
Specifically, the HAP could adaptively schedule the transmission scheme from WET, OMA, NOMA, and WET+OMA, optimizing power allocation in the process.
Furthermore, we reformulated the EWSAoI minimization problem as a MDP and derived an optimal policy.
This policy, based on instantaneous AoI and remaining battery power, dictated scheme selection and power allocation. 
Numerical results substantiate the effectiveness of the proposed policy, with the optimal policy showcasing a distinctive decision boundary-switching property.

\section*{Acknowledgment}
This work was supported in part by the National Natural Science Foundation of China under grant No. 61901180 and 62101369, Natural Science Foundation of Guangdong Province under grant No. 2022A1515010111 and 2023A1515012890, Science $\&$ Technology Project of Guangzhou under grant No. 202201010574, Science and Technology Development Fund, Macau, SAR, under grant 0119/2020/A3, by University of Macau under Project Nos. MYRG2022-00242-FST and MYRG-GRG2023-00200-FST.

\bibliographystyle{ieeetr}
\bibliography{ref}

\end{document}